\documentclass[jkps,preprint,fleqn,showpacs,showkeys]{revtex4}

\begin{document}

\setcounter{page}{1}
\title{On Shell Renormalization Scheme From the Loopwise Expansion of the Pole Mass}
\author{Chungku Kim}
\email{kimck@kmu.ac.kr}
\thanks{Fax: +82-053-580-5219}
\affiliation{Department of Physics, Keimyung University, Daegu 42601, Korea}

\date{\today}

\begin{abstract}
We introduce an on shell renormalization scheme in which the
mass parameter of minimal MS scheme is replaced with the pole mass obtained
from the loop order expansion of the pole mass in the MS scheme. As a consequence,
the quartic coupling constant remains same as that of the MS scheme and the
vacuum expectation value gets contributions from the one-particle-irreducible
diagrams. We also show the renormalization group invariance of the pole mass in this scheme.
\end{abstract}

\maketitle


The pole mass plays an important role in the process where the
characteristic scale is close to the mass shell\cite{Narison-1}. It was
shown that the pole mass is infrared finite and gauge invariant\cite
{Kronfeld} and is also invariant under the renormalization group(RG)\cite
{Kim}. The relation between the bare mass $M_{B}^{2}$ and the pole mass $%
M^{2}$ in the on shell scheme\cite{on-shell} is given by 
\begin{equation}
M_{B}^{2}=M^{2}+\delta M^{2}.
\end{equation}
The mass counterterm $\delta M^{2}$ is given by 
\begin{equation}
\delta M^{2}=-\left[ \Pi (p^{2},m_{B}^{2},\text{ }\lambda _{B})\right]
_{p^{2}=-M^{2}},
\end{equation}
in Euclidean space-time. Here the self energy $\Pi (p^{2},m_{B}^{2},$ $%
\lambda _{B})$ is either given by the one-particle-irreducible diagrams
where the contributions from the vacuum expectation value(VEV) $v^{2}\ $are
included in the pole mass as in \cite{Sirlin} or given by the sum of the
one-particle-irreducible and one-particle-reducible diagrams as in [6,7]
where only the contributions from the tree level vacuum expectation values $%
v_{0}^{2}$ are included in the pole mass and those from the higher order
vacuum expectation values such as $\lambda v_{0}v_{1}+\frac{\lambda }{2}%
v_{1}^{2}+\cdots $ contribute one-particle-reducible diagrams to $\Pi
(p^{2},m_{B}^{2},$ $\lambda _{B})$. However, the RG running of the Higgs
quartic coupling constant which is used in the vacuum stability analysis 
\cite{stability} is given in MS renormalization scheme\cite{beta} whereas in
the former case, the Higgs quartic coupling constant does not coincide with
that of minimal subtraction(MS) renormalization scheme. Hence, 
although the RG running of the Higgs coupling constant appear to coincide at
 one loop, the coincidence to all orders is not guaranteed
and needs further investigation. The latter renormalization scheme, known as
Fleischer and Jegerlehner(FJ) scheme, is widely used in the two-Higgs-doublet
models[10,11] recently. In this paper, we will introduce a procedure for an
on shell renormalization scheme in which the mass parameter of the MS scheme is
replaced with the pole mass obtained from the loop order expansion of the
pole mass in the MS scheme. In order to do this, we first obtain the pole mass
as a function of mass and the Higgs quartic coupling constant of the MS scheme
in a loop order. Then, by inverting this series, we obtain the mass parameter
of the MS scheme as a function of the pole mass and the Higgs quartic coupling
constant of the MS scheme. It turns out that the resulting vacuum expectation
value(VEV) contains not only the tadpole diagrams but the 
one-particle-irreducible(1PI) self energy
diagrams. Since this is a finite transformation between the mass parameters,
the renormalization constant of the Higgs quartic coupling constant remains same as
the one in the MS scheme and hence have the same RG running. 

The bare Lagrangian for the neutral scalar field theory with spontaneous
symmetry breaking in the Euclidean space-time is given by 
\begin{equation}
L_{B}=\frac{1}{2}(\partial \Phi _{B})^{2}-\frac{1}{2}m_{B}^{2}(\Phi
_{B}+v_{B})^{2}+\frac{1}{24}\lambda _{B}(\Phi _{B}+v_{B})^{4}.
\end{equation}
The VEV is obtained from the minimum condition for
the renormalized effective potential $V_{eff}(\phi )$ where $\phi $ is the
classical field: 
\begin{equation}
0=\frac{1}{v_{B}}\left[ \frac{\partial V_{eff}(\phi )}{\partial \phi }%
\right] _{\phi =0}=-m_{B}^{2}+\frac{1}{6}\lambda _{B}v_{B}^{2}+\Pi
^{tad}(m_{B}^{2},\text{ }\lambda _{B}).
\end{equation}
where $\Pi^{tad}(m_{B}^{2},\text{ }\lambda _{B})$ is the unrenormalized tadpole.
The bare quantities are related to the renormalized quantities as 
\begin{eqnarray}
&&\Phi _{B}=\sqrt{Z_{\phi }}\Phi =\sqrt{1+\delta Z_{2}+\cdots }\Phi ,\text{ }%
v_{B}=\sqrt{Z_{v}}v=\sqrt{1+\delta v_{2}+\cdots }v, 
\nonumber \\
&&m_{B}^{2}=m^{2}(1+\delta m_{1}^{2}+\cdots )\text{ and }\lambda _{B}=\lambda 
\text{ }(1+\delta \lambda _{1}+\cdots ),
\end{eqnarray}
where we have used the fact that $\Phi _{B}$ and $v_{B}$ have same
renormalization constants which vanishes at one-loop in neutral scalar
theory($\delta Z_{1}=0$). By solving Eq.(4), we obtain the VEV as a series
in the loop order expansion 
\begin{equation}
v=v_{0}+v_{1}+v_{2}+\cdots \text{ }(\text{ }v_{0}^{2}=\frac{6m^{2}}{\lambda }%
\text{ }).
\end{equation}
The tree level relation between the MS mass $m$ and the pole mass $M$ is
given by 
\begin{equation}
M^{2}=-m^{2}+\frac{1}{2}\text{ }\lambda v_{0}^{2}=2m^{2},
\end{equation}
Then, in order to obtain the one-loop counterterms in the broken symmetric
phase, let us consider the one-loop effective potential\cite{EP} \ including
the one-loop terms obtained by substituting Eqs.(5) and (6) into Eq.(3) as 
\begin{eqnarray}
&&\frac{1}{2}\int \frac{d^{D}p}{(2\pi )^{D}}\log (p^{2}+2m^{2}+\lambda
v_{0}\phi +\frac{1}{2}\lambda \phi ^{2})+\{2m^{2}v_{1}+v_{0}m^{2}(-\delta
m_{1}^{2}+\delta \lambda _{1})\}\phi 
\nonumber \\
&&+\frac{1}{2}\{\lambda
v_{0}v_{1}+m^{2}(-\delta m_{1}^{2}+3\delta \lambda _{1})\}\phi ^{2} 
+\frac{1}{6}(\lambda v_{1}+\delta \lambda _{1}v_{0})\phi ^{3}+\frac{1}{24}%
\delta \lambda _{1}\phi ^{4}  \nonumber \\
&=&-\frac{1}{4}(2m^{2}+\lambda v_{0}\phi +\frac{1}{2}\lambda \phi ^{2})^{2}(%
\frac{1}{\varepsilon }-Log(\frac{2m^{2}+\lambda v_{0}\phi +\frac{1}{2}%
\lambda \phi ^{2}}{\mu ^{2}}+\frac{3}{2})  \nonumber \\
&&+\{2m^{2}v_{1}+v_{0}m^{2}(-\delta m_{1}^{2}+\delta \lambda _{1})\}\phi
+\frac{1}{2}\{\lambda v_{0}v_{1}+m^{2}(-\delta m_{1}^{2}+3\delta \lambda
_{1})\}\phi ^{2}  \nonumber \\
&& +\frac{1}{6}(\lambda v_{1}+\delta \lambda _{1}v_{0})\phi
^{3}+\frac{1}{24}\delta \lambda _{1}\phi ^{4},
\end{eqnarray}
where we have used the $D=4-2\varepsilon $ dimensional regularization. By
noting that the one-loop renormalization constant of $Z_{\phi }$ is
zero in the neutral scalar theory and using the one-loop counterterms in
symmetric phase given by\cite{Ramond}\ 
\begin{equation}
\delta m_{1}^{2}=\frac{\lambda }{2\varepsilon }m^{2}\text{ and \ }\delta
\lambda _{1}=\frac{3}{2\varepsilon }\lambda .
\end{equation}
we can check that the $\frac{1}{\varepsilon }$ poles of the effective potential
in the broken symmetry phase given in Eq.(8) can be removed by the one-loop
counterterms of the symmetric phase in
the loop order expansion. The one-loop VEV term $v_{1}$ can be obtained from the
vanishing condition of the one-point function.
Now, let us introduce a procedure in which the mass parameter of minimal
subtraction(MS) scheme is replaced with the pole mass obtained from the loop
order expansion pole mass in the MS scheme. The pole mass $M^{2}$ is
defined as the pole of the renormalized inverse two point function $\Gamma
^{1PI}(p^{2},m^{2},$ $\lambda )$ obtained from the unrenormalized
1PI self-energy $\Pi^{1PI}(p^{2},m_{B}^{2},$ 
$\lambda _{B})$ as 
\begin{equation}
\Gamma ^{1PI}(p^{2})=Z_{\phi }\text{ }[\text{ }p^{2}-m_{B}^{2}+\frac{\lambda
_{B}}{2}\text{ }v_{B}^{2}+\Pi ^{1PI}(p^{2},m_{B}^{2},\text{ }\lambda _{B})%
\text{ }]=0\text{ when }p^{2}=-M^{2}.
\end{equation}
The tree level bare mass term of Eq.(3) becomes the bare mass in the on shell
scheme as
\begin{equation}
M_{B}^{2}=M^{2}+\delta M^{2}=-m_{B}^{2}+\frac{\lambda _{B}}{2}\text{ }%
v_{B}^{2},
\end{equation}
and hence Eq.(10) gives the renormalization condition for $\delta M^{2}$ as 
\begin{equation}
\delta M^{2}+\left[ \Pi ^{1PI}(p^{2},M_{B}^{2},\text{ }\lambda _{B})\right]
_{p^{2}=-M^{2}}=0.
\end{equation}
We can check that the $\frac{1}{\varepsilon }$ pole the of the one-loop
two-point function in the broken symmetric phase given by 
\begin{eqnarray}
\Pi ^{1PI}(p^{2},m^{2},\text{ }\lambda ) &=&
\begin{picture}(50,20) \put(8,-4){\text{..........}}
\put(20,5){\circle{16}}\end{picture} +\begin{picture}(50,20)
\put(8,4){\text{....}} \put(34,4){\text{....}} \put(26,5){\circle{16}}
\end{picture}  \nonumber \\
&=&\frac{\lambda }{2}A(2m^{2})-\frac{1}{2}\lambda
^{2}v_{0}^{2}B(p^{2},2m^{2}),
\end{eqnarray}
can be removed by using the one-loop counterterms in the symmetric phase
given by $\lambda v_{0}v_{1}+m^{2}\text{ }(-\delta m_{1}^{2}+3\delta \lambda _{1})$
(see Eqs.(8) and (9)) in the loop order expansion. 
Here the one-loop function $A(m^{2})$\ and $B(p^{2},m^{2})$\ is given by%
\cite{Passarino}\ 
\begin{equation}
A(m^{2})=\int \frac{d^{D}q}{(2\pi )^{D}}\frac{1}{q^{2}+m^{2}},
\end{equation}
and 
\begin{equation}
B(p^{2},m^{2})=\int \frac{d^{D}q}{(2\pi )^{D}}\frac{1}{%
(q^{2}+m^{2})((p+q)^{2}+m^{2})},
\end{equation}
Then we can obtain the loop order expansion of the pole mass $M^{2}$ as 
\begin{eqnarray}
M^{2} &=&2m^{2}+\lambda \text{ }v_{0}\text{ }v_{1}+\Pi
_{1}^{1PI}(-2m^{2},m^{2},\text{ }\lambda )+\left[ \frac{d\Pi
_{1}^{1PI}(p^{2},m^{2},\text{ }\lambda )}{dp^{2}}\right]
_{p^{2}=-2m^{2}}(\lambda \text{ }v_{0}\text{ }v_{1}  \nonumber \\
&&+\Pi
_{1}^{1PI}(-2m^{2},m^{2},\text{ }\lambda ))
+\Pi _{2}^{1PI}(-2m^{2},m^{2},\text{ }\lambda )+\lambda \text{ }v_{0}\text{
}v_{2}+\frac{\lambda }{2}\text{ }v_{1}^{2}+\cdots ,
\end{eqnarray}
where $\Pi _{l}^{1PI}(p^{2},m^{2},$ $\lambda )$ \ is the renormalized $l$%
-loop 1PI self-energy obtained in the MS scheme. By solving Eq.(4), we can
obtain $v_{i}$ as a function of $m^{2}$ and $\lambda $ and hence Eq.(16)
determines the pole mass $M^{2}$ as function of $m^{2}$ and $\lambda $. At
tree level, we obtain the mass relation $M^{2}=2m^{2}$ as in Eq.(7). Then,
by inverting Eq.(18), we can obtain the loop order expansion of $m^{2}$ as
function of $M^{2}$ as 
\begin{equation}
m^{2}=\frac{1}{2}M^{2}-M_{1}^{2}(M^{2})-M_{2}^{2}(M^{2})-\cdots ,
\end{equation}
where $M_{l}^{2}(M^{2})$ is the $l-$loop terms in the expansion. Moreover,
if we write the series expansion of the VEV\ where the pole mass $M^{2}$ is
the tree level mass parameter as in Eq.(17), the order of the VEV changes
also. For example, $v_{0}$ given in Eq.(6) becomes infinite series of
function of $M^{2}$ which consists of not only tadpole diagrams but 1PI \
diagrams 
\begin{equation}
v_{0}(m^{2})=\sqrt{\frac{6m^{2}}{\lambda }}\simeq \sqrt{\frac{3M^{2}}{%
\lambda }}(1-\frac{M_{1}^{2}}{M^{2}}+\cdots ),
\end{equation}
and let us write the resulting series expansion of the VEV\ where the pole
mass $M^{2}$ is the tree level mass parameter as 
\begin{equation}
v=\overline{v_{0}}+\overline{v_{1}}+\overline{v_{2}}+\cdots \text{ ,}
\end{equation}
where the relation between $v$ and $\overline{v}$ is 
\begin{equation}
\text{ }\overline{v_{0}}=\left[ v_{0}\right] _{m^{2}=\frac{1}{2}M^{2}}=\sqrt{%
\frac{3M^{2}}{\lambda }\text{ }}\text{and }\overline{v_{1}}=-\frac{M_{1}^{2}%
}{M^{2}}\text{ }\overline{v_{0}}\text{ }+\left[ v_{1}\right] _{m^{2}=\frac{1%
}{2}M^{2}}\text{\ etc.}
\end{equation}
Now, $M_{l}^{2}(M^{2})$ and $\overline{v_{l}}$ can be determined if we
substitute Eqs.(17) and (19) into the two conditions given in Eqs.(4) and
(15). At one-loop, Eqs.(4) and (10) gives 
\begin{equation}
\text{ }-\frac{1}{2}M^{2}\text{ }\delta m_{1}^{2}\text{ }+M_{1}^{2}+\frac{%
\text{ }\delta \lambda _{1}}{6}\text{ }\overline{v_{0}}\text{ }\overline{%
v_{1}}\text{ }+\frac{\lambda }{3}\text{ }\overline{v_{0}}\text{ }\overline{%
v_{1}}+\text{ }\Pi _{1}^{tad}(m^{2},\text{ }\lambda )=0,
\end{equation}
and 
\begin{equation}
-\frac{1}{2}M^{2}\text{ }\delta m_{1}^{2}+M_{1}^{2}+\frac{\delta \lambda _{1}%
}{2}\text{ }\overline{v_{0}}^{2}+\lambda \text{ }\overline{v_{0}}\text{ }%
\overline{v_{1}}+\text{ }\Pi _{1}^{1PI}(-M^{2},m^{2},\text{ }\lambda )=0,
\end{equation}
where $\delta m_{1}^{2}$ and $\delta \lambda _{1}$ is the one-loop
counterterms for $m^{2}$\ and $\lambda $ and\ we have used the fact that $%
Z_{\phi }=1$ up to one-loop order. Since Eq.(17) is a finite transformation
of the mass parameters from $m^{2}$ to the pole mass $M^{2}$, all the $\frac{%
1}{\varepsilon }$ poles will be removed by the renormalization constants of
the MS scheme. Actually, by using $\Pi _{1}^{1PI}(-M^{2})$ given in Eq.(13)
and the one-loop renormalization constants for the neutral scalar
theory in the MS scheme given in Eq.(9),
we can see that the $\frac{1}{\varepsilon }$ pole in the one-loop functions
vanishes in Eqs. (21) and (22) as expected. Then, by solving the remaining
finite equations, we can determine $M_{1}^{2}$ and $\overline{v_{1}}$ as 
\begin{equation}
M_{1}^{2}\text{ }=\frac{1}{2}\text{ }\Pi _{1}^{1PI}(-M^{2},M^{2},\text{ }%
\lambda )-\frac{3}{2}\text{ }\Pi _{1}^{tad}(M^{2},\text{ }\lambda )=\left[ -%
\frac{1}{2}\lambda \text{ }A(M^{2})-\frac{3}{4}M^{2}B(-M^{2},M^{2})\right]
_{finite},
\end{equation}
and 
\begin{equation}
\lambda \text{ }\overline{v_{0}}\text{ }\overline{v_{1}}=-\frac{3}{2}\text{ }%
\Pi _{1}^{1PI}(-M^{2},M^{2},\text{ }\lambda )+\frac{3}{2}\text{ }\Pi
_{1}^{tad}(M^{2},\text{ }\lambda )=\left[ \frac{9}{4}M^{2}B(-M^{2},M^{2})%
\right] _{finite},
\end{equation}
where we have used Eq.(13) and the renormalized one-loop tadpole $\Pi
_{1}^{tad}(p^{2},m^{2},$ $\lambda )$\ which can be obtained from Eq.(4) as 
\begin{equation}
\Pi _{1}^{tad}(M^{2},\text{ }\lambda )=\frac{1}{2}\lambda A(M^{2}),
\end{equation}
at one-loop order. $\left[ X\right] _{finite}$ means the finite part of $X$ $%
\ $obtained by removing the $\frac{1}{\varepsilon }$ pole of $X.$\ 
Now let us consider the RG invariance of the pole mass in this scheme. Since
the bare mass $M_{B}^{2}$ is RG invariant, we can see from Eq.(11) that the
RG invariance of the pole mass requires the RG invariance of $\delta M^{2}$
so that
\begin{equation}
\mu \frac{d}{d\mu }\delta M^{2}=(\mu \frac{\partial }{\partial \mu }+\beta
_{\lambda }\frac{\partial }{\partial \lambda }+\beta _{m^{2}}\frac{\partial 
}{\partial m^{2}})\delta M^{2}=0.
\end{equation}
By substituting Eq.(13) into (12), we obtain one-loop mass counterterm in the
on shell scheme as
\begin{equation}
\delta M_{1}^{2}=-\frac{\lambda }{2}A(M^{2})+\frac{3 \lambda}{2}
M^{2}B(- M^{2},M^{2})
\end{equation} and by noting that RG function $\beta _{\lambda }$ is given by 
\begin{equation}
\beta _{\lambda }=-2\varepsilon \lambda +3\lambda ^{2}+\cdot \cdot \cdot ,
\end{equation}
we can see that the one loop counterterm given in Eq.(28) satisfies Eq.(26)
and hence pole mass is RG invariant up to one loop. 
The VEV can be obtained directly from Eqs.(4),(11) and (12). In order to see
this, let us eliminate $m_{B}^{2}$ in Eq.(4) by using Eq.(11) to obtain 
\begin{equation}
M^{2}+\delta M^{2}-\frac{1}{3}\text{ }\lambda _{B}v_{B}^{2}+\Pi
^{tad}(M_{B}^{2},\text{ }\lambda _{B})=0,
\end{equation}
and then, by using Eq.(12) we obtain 
\begin{equation}
\text{ }\lambda _{B}v_{B}^{2}=3[M^{2}-\Pi^{1PI}(p^{2},M_{B}^{2},\text{ 
}\lambda _{B})+\Pi^{tad}(M_{B}^{2},\text{ }\lambda _{B})].
\end{equation}
After eliminating $\frac{1}{\varepsilon }$ pole with the MS renormalization
constants, we obtain the formula for the renormalized VEV as 
\begin{equation}
\lambda v^{2}=3\left[M^{2}-\text{ }\Pi ^{1PI}(-M^{2},M^{2},\text{ }\lambda )+\Pi
^{tad}(M^{2},\text{ }\lambda )\right] _{finite},
\end{equation}
which agrees with the results given in Eqs.(20) and (24). In this way, we
can determine $\delta M^{2}$ and $\overline{v}$ from Eqs.(12) and (29)
without need to calculate $M_{l}^{2}(M^{2})$.

In this paper, we have introduced a procedure for an on shell
renormalization scheme in which the mass parameter of minimal MS scheme is
replaced with the pole mass obtained from the loop order expansion of
the pole mass in the MS scheme. In order to see the difference between the on
shell renormalization scheme based on the loopwise expansion introduced in
 this paper and previously known schemes, let us consider the case of
the Sirlin and Zucchini(SZ) scheme\cite{Sirlin} which is used most frequently. 
First, in SZ scheme, the Higgs mass $M_{H}$ is related to running mass $m$ as
\begin{equation}
M_{H}^{2}=\frac{1}{3}\lambda _{Sirlin}v_{0}^{2}=2m^{2}
\end{equation}
Note the different normalization for $\lambda$ between Ref.[5] and our paper. 
In our scheme, the Higgs pole mass is obtained
 by loopwise inversion of the defining equation of the Higgs pole mass given in 
Eq.(10) and as a result the corresponding relation becomes
\begin{equation}
M_{H}^{2}-2M_{1}^{2}-2M_{1}^{2}-\cdots =\frac{1}{3}\lambda _{MS}v_{0}^{2}=2m^{2}
\end{equation}
as in Eq.(16). By noting that $\beta _{m^{2}}=\lambda m^{2}$ and by using
$M_{1}^{2}$ given in Eq.(23), we can see that both sides of Eq.(33) have same RG
running. If we extend the on shell renormalization scheme based on loopwise
 expansion to the electroweak sector of the standard model, we can choose the 
parameter set as  $\left\{ {G_\mu, M_H^2, M_W^2, \lambda _{MS}}\right\} $
 instead of $\left\{ {G_\mu, M_H^2, M_W^2, \lambda _{Sirlin}}\right\} $ so that the RG
evolution of the Higgs quartic coupling constant $\lambda$ which is important to determine 
the vacuum stability condition can be obtained by the RG functions of the MS scheme.
Second, the vacuum expectation value $v$ that emerges at the triple scalar vertex
 including the Higgs (HHH, HGG etc.)  have a loopwise expansion and should be determined 
order by order from Eq.(30). This equation shows that the vacuum expectation value gets
contributions not only from the tadpoles as in previous schemes 
but also from the one-particle-irreducible diagrams in the loopwise expansion scheme.
We have investigated in the neutral scalar field theory which is most simple model and
the extension to more complicated models like Standard Model is under investigation.

\end{document}